\documentclass[11pt]{article}
\usepackage{rotating}
\usepackage{amsbsy}
\usepackage{epsfig}
\usepackage{subfig}
\usepackage{url}
\usepackage{color}
\usepackage{multirow}
\usepackage{amsmath}
\usepackage{amsfonts}
\usepackage{array}
\usepackage[pagebackref]{hyperref}
\usepackage{authblk}

\newcommand{\sym}[1]{{\sf #1}}

\newcommand{\len}{\mbox{{\sf len}}}

\newcommand{\encrypt}{{\sf Encrypt}}

\newcommand{\poly}{{\sf Horner}}
\newcommand{\feistel}{{\sf Feistel}}

\newcommand{\rF}{\mbox{$\mathbb{F}$}}

\def\ArtWork#1{\noindent\hfill\epsfbox{#1}\hfill}%

{\bfseries}{\itshape}

\begin{document}
\title{Fast Low Level Disk Encryption Using FPGAs}

\author[1]{Debrup Chakraborty}
\author[2]{Sebati Ghosh}
\author[3]{Cuauhtemoc Mancillas L\'opez}
\author[1]{Palash Sarkar}
\affil[1]{Indian Statistical Institute, 203, B.T. Road, Kolkata, India 700108}
\affil[2]{Department of Computer Science, University of York, Deramore Lane, York YO10 5GH}
\affil[3]{Computer Science Department, CINVESTAV-IPN, Mexico, D.F., 07360, Mexico}
\affil[ ]{Emails: {\tt debrup@isical.ac.in, sebati1987@gmail.com, cuauhtemoc.mancillas83@gmail.com, palash@isical.ac.in}}

\date{}
\maketitle
\newcommand{\RAND}{\mbox{{\scriptsize {\sc RAND}}}}
\newcommand{\HCH}{\mbox{{\scriptsize {\sc HCH}}}}
\newcommand{\NON}{\mbox{{\scriptsize {\sc NON}}}}
\newcommand{\pf}{{\bf Proof. }}
\begin{abstract}
	A fixed length tweakable enciphering scheme (TES) is the appropriate cryptographic functionality for low level disk encryption. Research on TES over the
	last two decades have led to a number of proposals many of which have already been implemented using FPGAs. This paper considers the FPGA implementations of
	two more recent and promising TESs, namely AEZ and FAST. The relevant architectures are described and simulation results on the Xilinx Virtex 5 and Virtex 7
	FPGAs are presented. For comparison, two IEEE standard schemes, XCB and EME2 are considered. The results indicate that FAST outperforms the other schemes
	making it a serious candidate for future incorporation by disk manufacturers and standardisation bodies. 
\end{abstract}

\section{Introduction\label{sec-intro}}
Sensitive data reside on hard disks of computers. It is important to be able to protect such data from unauthorised access and tampering. This requires an appropriate
mechanism for keeping the data on the disk in an encrypted form. High level applications, on the other hand, operate on unencrypted data. A disk encryption module
performs the task of converting to and from encrypted data. While the primary requirement for a disk encryption algorithm is to ensure the relevant notion of security,
it is also important to ensure that there is no substantial performance degradation due to storing the data in encrypted form. To a large extent, the performance of
a disk encryption algorithm depends at the level where it is positioned. For example, a software based disk encryption algorithm will perform the encryption and
decryption operations in software. Consequently, there will be a loss of efficiency. Low level disk encryption, on the other hand, envisages the disk encryption
module to be placed just above the disk controller and requires the module to be implemented in hardware. If the speed of encryption/decryption matches the speed
of read and write operations on the disk, then there will be no performance penalty due to use of an encrypted disk. 

The present work considers low level disk encryption implemented in hardware. 
A disk is organised into sectors. In modern disks, a sector stores 4096 bytes. Each sector has a sector address. 
The cryptographic primitive which is well suited for disk encryption is tweakable
enciphering scheme (TES)~\cite{DBLP:conf/crypto/HaleviR03}. The possibility of using a deterministic authenticated encryption with associated data (DAEAD) scheme~\cite{DBLP:conf/eurocrypt/RogawayS06} for disk encryption has been 
considered in~\cite{DBLP:journals/jce/ChakrabortyMS18}. Doing this would require disk manufacturers to allocate a nominal extra storage per sector to accommodate the ciphertext produced by 
a DAEAD scheme which is a few bytes longer than the message. As has been demonstrated in~\cite{DBLP:journals/jce/ChakrabortyMS18}, disk encryption based on a DAEAD scheme can be more efficient
than one based on a TES. Nonetheless, until the time disk manufacturers actually decide to alter the designs of physical disk sectors, it is TES which has to be considered
for disk encryption.

A TES is used
to perform sectorwise encryption/decryption. The encryption algorithm of a TES uses a secret key $K$, a tweak $T$, and a message to produce a ciphertext which is of the
same length as the message; the decryption algorithm uses $K$, $T$ and the ciphertext to produce a message which is of the same length as the ciphertext. In the 
disk encryption application, the tweak is taken to be the sector address and the message is the data that is to be written to the sector. The ciphertext produced by the
TES encryption algorithm is physically written to the sector. The actual message is not stored anywhere. Decryption works by reversing the steps, i.e., the physical content
of the sector is decrypted using the sector address as the tweak and the message obtained is returned to the high level application. It is to be noted that since the clear
message is not stored anywhere, if the secret key $K$ is lost, then the disk becomes unreadable. 

Starting from the pioneering work of Halevi and Rogaway~\cite{DBLP:conf/crypto/HaleviR03}, over the years a number of TESs suitable for disk encryption have been 
proposed~\cite{DBLP:conf/ctrsa/HaleviR04,DBLP:conf/indocrypt/Halevi04,cryptoeprint:2004:278,DBLP:conf/sacrypt/McGrewF07,DBLP:conf/cisc/WangFW05,DBLP:conf/indocrypt/ChakrabortyS06,DBLP:conf/crypto/Halevi07,DBLP:conf/icisc/Sarkar07,Sa09,Sa11-IPL,DBLP:journals/tc/ChakrabortyMS15,DBLP:conf/eurocrypt/HoangKR15,DBLP:conf/asiacrypt/BhaumikN15,FAST,DBLP:journals/tosc/CrowleyB18}.
Out of these, CMC~\cite{DBLP:conf/crypto/HaleviR03}, EME~\cite{DBLP:conf/ctrsa/HaleviR04} (and its variants~\cite{DBLP:conf/indocrypt/Halevi04}) and 
AEZ~\cite{DBLP:conf/eurocrypt/HoangKR15} use only a block cipher, while most of the other schemes use a block cipher 
along with a XOR universal hash function. IEEE has standardised two schemes~\cite{P161911}.
For implementation of a TES built using a block cipher, the instantiation of the block cipher is typically done using the standardised advanced encryption system (AES). 
A suite of stream cipher and lightweight hash function
based TES has been presented in~\cite{DBLP:journals/tc/ChakrabortyMS15} which is suitable for low area, low power and low cost applications of disk encryption. Another TES proposal called
Adiantum~\cite{DBLP:journals/tosc/CrowleyB18} is built from AES, the stream cipher ChaCha, along with two hash functions. Adiantum is targeted towards software implementation on low end processors 
which do not have intrinsic processor support for AES and 64-bit polynomial multiplication. 

Most of the block cipher based TES proposals require both the encryption and the decryption algorithms of the block cipher. So, hardware implementations require
the implementations of both the encryption and the decryption modules of the underlying block cipher. For one thing, this increases the area of the hardware.
In the case of AES, the critical path of the decryption module of AES is longer than that of the encryption module. So, the requirement of implementing both the
encryption and the decryption modules also leads to an increase of the critical path length. 

Presently, there are three known TES proposals, namely AEZ~\cite{DBLP:conf/eurocrypt/HoangKR15}, FAST~\cite{FAST} (also its predecessor~\cite{Sa11-IPL}) and 
FMix~\cite{DBLP:conf/asiacrypt/BhaumikN15}, which use only the encryption module
of the underlying block cipher, i.e., both the encryption and the decryption algorithms of these TESs are built using only the encryption function of the block cipher.
Consequently, hardware implementation of such TESs using AES will require smaller area as well as a smaller critical path. Of these three TESs, FMix is 
a sequential construction, i.e., the blocks of the message are processed in a sequential manner. Such a design cannot profit from the various pipelining and parallelism
options that can be implemented in hardware. In view of this, we do not consider FMix for implementation. 

The main contribution of the present work is to present efficient implementations of FAST~\cite{FAST} and AEZ~\cite{DBLP:conf/eurocrypt/HoangKR15} on FPGAs. Our target is to
obtain high speed, so we do not consider low power and low area, but slower FPGAs. While our implementations are on FPGA, in actual deployment ASICs will be used
leading to even higher speed. As explained above, our rationale for choosing FAST and AEZ for implementation is that presently these are the only two TESs which are parallelisable
and are built using only the encryption function of the underlying block cipher. For comparison, we consider previous implementations of the IEEE standards
XCB and EME2. Of the four, it turns out that FAST provides the highest throughput and also the smallest area. 
This makes FAST an attractive option to be considered for deployment and standardisation. 

Previous FPGA implementations of TESs have been reported in~\cite{DBLP:journals/tc/Mancillas-LopezCR10,DBLP:journals/tc/ChakrabortyMRS13}. The work~\cite{DBLP:journals/tc/Mancillas-LopezCR10} comprehensively implemented all TESs proposed prior to 2010, while
the work~\cite{DBLP:journals/tc/ChakrabortyMRS13}, published in 2013, considered the TESs proposed in the intervening period. The present work may be seen as a continuation of 
the prior work on FPGA implementations of TESs and brings the literature up to date on this topic. 

A pipelined architecture for the AES encryption function based on ideas from~\cite{DBLP:conf/africacrypt/BulensSQPR08} has been made. This is required for both FAST and AEZ.
FAST provides two options for the hash function, either
a Horner based computation, or, a hash function based on the Bernstein-Rabin-Winograd (BRW) polynomials. The implementation of the Horner based hash function
uses a decimated approach and utilises two multipliers. The BRW-based hash function design is for 255-block messages using a 4-stage Karatsuba multiplier. The
only previously known BRW implementation in hardware was for 31-block messages using a 3-stage multiplier~\cite{DBLP:journals/tc/ChakrabortyMRS13}. 
There has been only one previous work on the hardware implementation of AEZ~\cite{DBLP:conf/indocrypt/HomsirikamolG16}. This work used an iterated round implementation and had indicated that AEZ
is difficult to implement in hardware. Our hardware for AEZ uses a pipelined implementation. A troublesome issue in AEZ implementation is the computation of
the internal masking values. We implement two approaches to performing such computation, namely one is on the fly and the other is to pre-compute and store them in the
memory. Overall, our designs for FAST and AEZ present novel ideas which can turn out to be useful in other contexts.

In Section~\ref{sec-prelim} we present the notation and the descriptions of FAST and AEZ. The description is tailored to the requirement of disk encryption.
For further details including formal security analysis, we refer the reader to~\cite{FAST} for FAST and to~\cite{DBLP:conf/eurocrypt/HoangKR15} for AEZ.
The implementation of FAST is described in Section~\ref{sec-FAST}, while
the implementation of AEZ is described in Section~\ref{sec-AEZ}. Results are compared in Section~\ref{sec-compare}. The paper is concluded in Section~\ref{sec-conclu}.

\section{Preliminaries\label{sec-prelim}}
For the description of the preliminaries, we fix a positive integer $n$. This represents the block size of the underlying block cipher. Since our actual
implementations are for the AES, for the implementations, we will use $n=128$.

\paragraph{\bf Notation:}
Let $X$ and $Y$ be binary strings.
\begin{description}
\item{$\bullet$} The length of $X$ will be denoted as $\len(X)$. 
\item{$\bullet$} The concatenation of $X$ and $Y$ will be denoted as $X||Y$. 
\item{$\bullet$} For an integer $i$ with $0\leq i<2^n$, $\sym{bin}_n(i)$ denotes the $n$-bit binary representation of $i$.
\item{$\bullet$} ${\sf parse}_n(Y)$: If $\len(Y)\geq 2n$, ${\sf parse}_n(Y)$ denotes $(Y_1,Y_2,Y_3)$ where
$\len(Y_1)=\len(Y_2)=n$ and $Y=Y_1||Y_2||Y_3$. In other words, ${\sf parse}_n(Y)$ divides the string $Y$ into three parts with the first
two parts having length $n$ bits each with the remaining bits of $Y$ (if any) forming the third part.
\end{description}

The description of FAST is in terms of a pseudo-random function (PRF) family $\{\mathbf{F}_{K}\}_{K\in\mathcal{K}}$, where for 
$K\in\mathcal{K}$, $\mathbf{F}_K:\{0,1\}^n\rightarrow\{0,1\}^n$. The concrete instantiation of $\mathbf{F}_K$ is done using the encryption function
of the AES which we denote as $E_K:\{0,1\}^n\rightarrow \{0,1\}^n$. AEZ is based on the encryption function of a block cipher and our implementation
instantiate the block cipher using $E_K$, the encryption function of AES.

\subsection{Description of FAST \label{subsec-FAST-desc}}
FAST is actually a suite of algorithms and can be customised to obtain various functionalities. In this paper, we provide the description that is required 
for disk encryption. In particular, this uses an $n$-bit tweak and a fixed length message. For the more general description of FAST, we refer to~\cite{FAST}.

FAST uses the PRF $\mathbf{F}$ to build a counter mode. Let $Y = Y_1||Y_2||\cdots||Y_m$ be a binary string, where $m\geq 1$ and each $Y_i$ is 
an $n$-bit string, $i=1,\ldots,m$. For $K\in\mathcal{K}$ and $S\in\{0,1\}^n$, we define $\sym{Ctr}_{K,S}(Y)$ in the following manner.
\begin{eqnarray}\label{eqn-Ctr}
{\sf Ctr}_{K,S}(Y)
& = & (S_1\oplus Y_1,\ldots,S_{m-1}\oplus Y_{m-1},S_m\oplus Y_m)
\end{eqnarray}
where $S_i=\mathbf{F}_K(S\oplus \sym{bin}_n(i))$. 

FAST requires a XOR universal hash function. Two options for the hash function have been proposed in~\cite{FAST}. Both of these hash functions
are defined over the finite field $\rF=GF(2^n)$. The field $\rF$ itself is represented using a fixed
primitive polynomial $\psi(\alpha)$ of degree $n$ over $GF(2)$. Under this representation, elements of $\rF$ can be considered to be $n$-bit binary strings.
The addition operation over $\rF$ will be denoted by $\oplus$;
for $X,Y \in \rF$, the product will be denoted as $XY$. The additive identity of $\rF$ will be denoted as $\mathbf{0}$ and will
be represented as $0^n$; the multiplicative identity of $\rF$ will be denoted as $\mathbf{1}$ and will be represented as $0^{n-1}1$.

Given an $n$-bit string $X$, it represents an element $X(\alpha)$ of $\rF$ represented using $\psi(\alpha)$. The operation $\alpha X(\alpha)\bmod \psi(\alpha)$
is the `multiply by $\alpha$' map and has been called a doubling operation~\cite{Ro04}.
For $n=128$, we use $\psi(\alpha)=\alpha^{128}\oplus\alpha^7\oplus\alpha^2\oplus\alpha\oplus 1$ to construct the field $\rF$.

Of the two hash functions used in FAST, one is based on evaluation of polynomials using Horner's rule and the other is based on the BRW polynomials. These are
defined below. 

\paragraph{\bf $\poly$:} For $m\geq 0$, let $\poly:\rF\times\rF^m\rightarrow\rF$ be defined as follows.
\begin{eqnarray*}
\poly(\tau,Y_1,\ldots,Y_m)
& = &
\left\{
\begin{array}{lcl}
\mathbf{0}, & & \mbox{if } m=0; \\
Y_1\tau^{m-1} \oplus Y_2\tau^{m-2} \oplus \cdots \oplus Y_{m-1}\tau \oplus Y_m, & & \mbox{if } m>0.
\end{array}\right.
\end{eqnarray*}
The notation $\poly_\tau(Y_1,\ldots,Y_m)$ denotes $\poly(\tau,Y_1,\ldots,Y_m)$.  

\paragraph{\bf BRW polynomials:}
For $m\geq 0$, let $\sym{BRW}:\rF\times\rF^m\rightarrow\rF$ be defined as follows.
\begin{tabbing}
\ \ \ \ \= $\bullet$ \= \ \ \ \ \=\ \ \ \ \= \kill
 \> $\bullet$ \> $\sym{BRW}_{\tau}() = \mathbf{0}$; \\
 \> $\bullet$ \> $\sym{BRW}_{\tau}(Y_1) = Y_1$; \\
 \> $\bullet$ \> $\sym{BRW}_{\tau}(Y_1,Y_2) = Y_1\tau\oplus Y_2$; \\
 \> $\bullet$ \> $\sym{BRW}_{\tau}(Y_1,Y_2,Y_3) = (\tau \oplus Y_1)(\tau^2 \oplus Y_2) \oplus Y_3$; \\
 \> $\bullet$ \> $\sym{BRW}_{\tau}(Y_1,Y_2, \cdots, Y_{m})$ \\
 \>           \> \> $= \sym{BRW}_{\tau}(Y_1,\cdots,Y_{t-1})(\tau^t \oplus Y_t)
		\oplus \sym{BRW}_{\tau}(Y_{t+1},\cdots,Y_{m})$; \\
 \> \> \> if  $t \in \{4,8,16,32,\cdots\}$ and $t\leq m < 2t$.
\end{tabbing}
We write $\sym{BRW}_{\tau}(\cdots)$ to denote $\sym{BRW}(\tau,\cdots)$. The important advantage of $\sym{BRW}$ over $\poly$ is that for 
$m\geq 3$, $\sym{BRW}_{\tau}(Y_1,\ldots,Y_{m})$ can be computed using $\lfloor m/2\rfloor$ field multiplications and
$\lfloor\lg m\rfloor$ additional field squarings to compute $\tau^2,\tau^4,\ldots$, whereas $\poly_{\tau}(Y_1,\ldots,Y_m)$ requires $m-1$ multiplications to
be evaluated.

We now provide the description of FAST for fixed length messages. 
Let $m\geq 3$ be an integer. We describe the encryption algorithm of FAST. The decryption algorithm can be derived from the encryption algorithm and is
presented in details in~\cite{FAST}. 

Consider an $m$-block message $X_1||X_2||\cdots||X_m$, where each $X_i$ is an $n$-bit block. For disk encryption application, the tweaks are sector addresses and we 
assume the tweak $T$ to be a single $n$-bit block. The encryption algorithm is shown in Table~\ref{tab-encrypt}. The sub-routine $\sym{Feistel}$ is shown in 
Table~\ref{tab-feistel}.
This encryption algorithm in Table~\ref{tab-encrypt} uses the functions $\mathbf{H}_{\tau}$ and
$\mathbf{G}_{\tau}^{\prime}$ which are built from two hash functions $h$ and $h^{\prime}$ in the following manner.
\begin{eqnarray}\label{eqn-H-Hprime}
\begin{array}{rcl}
\mathbf{H}_{\tau}(P_1,P_2,P_3,T) & = & (P_1\oplus h_{\tau}(T,P_3),P_2\oplus \tau (P_1\oplus h_{\tau}(T,P_3))); \\
\mathbf{G}_{\tau}^{\prime}(Y_1,Y_2,Y_3,T) & = & (Y_1\oplus \tau Y_2, Y_2\oplus h_{\tau}^{\prime}(T,Y_3)).
\end{array}
\end{eqnarray}

\begin{table*}
\begin{center}
\caption{\label{tab-encrypt}Encryption algorithm for {\sf FAST}.}
{\small
\begin{tabular}{|c|c|}
\hline
\epsfxsize=0.35\hsize \ArtWork{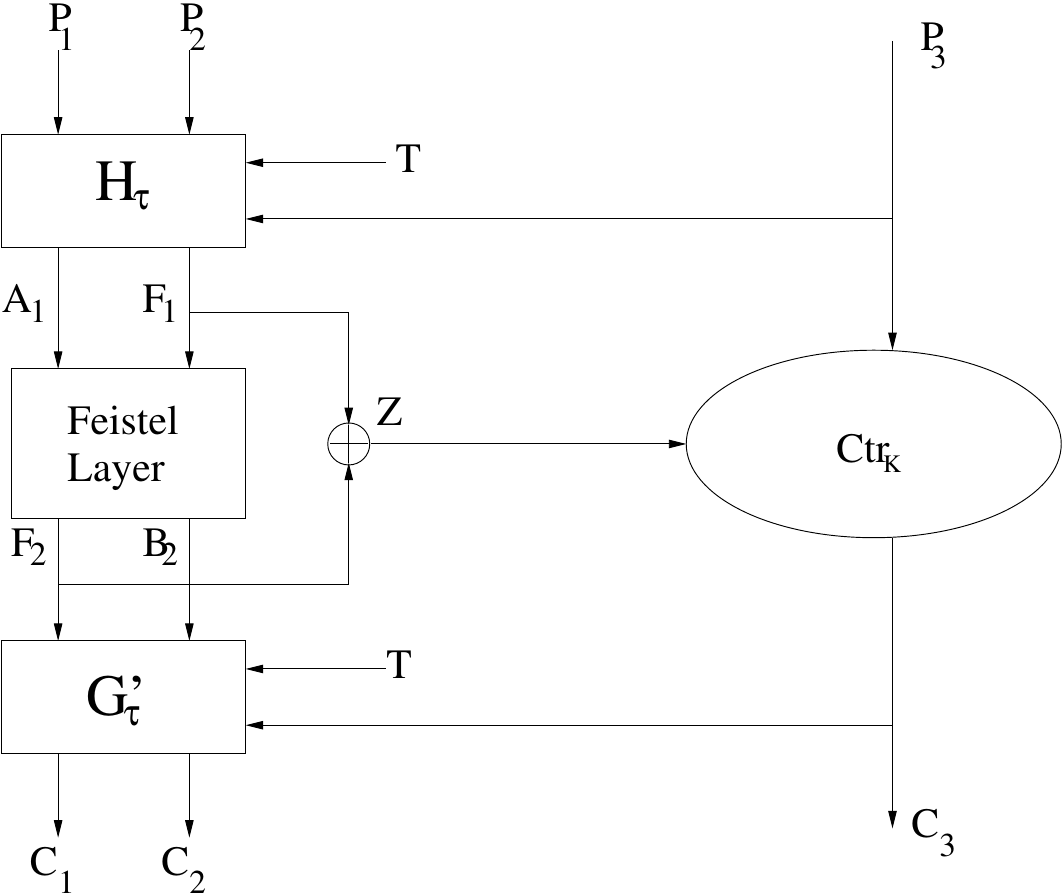}
&
\begin{minipage}[b]{210pt}
\begin{tabular}{l}
\begin{minipage}[b]{210pt}
\begin{tabbing}
\ \ \ \ \ \=\ \ \ \ \=\ \ \ \ \=\ \ \ \ \=\ \ \ \ \=\ \ \ \ \=\ \ \ \ \kill
{\bf Algorithm} ${\sf FAST}.{\encrypt}_{{K}}(T,X_1||X_2||\cdots||X_m)$\\
1.\> $\tau \leftarrow \mathbf{F}_K({\sf fStr})$;\\
2.\> $(X_1,X_2,X_3||\cdots||X_m) \leftarrow {\sf parse}_n(X_1||\cdots||X_m)$;\\
3.\> $(A_1,F_1)\leftarrow \mathbf{H}_{\tau}(X_1,X_2,X_3||\cdots||X_m,T)$; \\
4.\> $(F_2,B_2) \leftarrow \feistel_{K}(A_1,F_1)$;  \\
5.\> $Z \leftarrow F_1\oplus F_2$; \\
6.\> $C_3||\cdots||C_m \leftarrow \sym{Ctr}_{K,Z}(X_3||\cdots||X_m)$; \\
7.\> $(C_1,C_2)\leftarrow \mathbf{G}_{\tau}^{\prime}(F_2,B_2,C_3||\cdots||C_m,T)$; \\
8.\> return $(C_1||C_2||C_3||\cdots||C_m)$.
\end{tabbing}
\end{minipage}
\end{tabular} \\
\end{minipage}\\
\hline
\end{tabular}
}
\end{center}
\end{table*}
\begin{table*}
\begin{center}
\caption{\label{tab-feistel}A two-round Feistel construction required in Table~\ref{tab-encrypt}.}
{\small
\begin{tabular}{|m{100pt}|m{100pt}|}
\hline
\epsfxsize=0.13\vsize \ArtWork{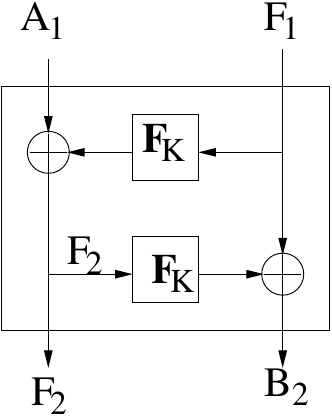}
&
\begin{minipage}[b]{220pt}
\begin{tabbing}
\ \ \ \ \ \=\ \ \ \ \=\ \ \ \ \=\ \ \ \ \=\ \ \ \ \=\ \ \ \ \=\ \ \
\ \kill
${\feistel}_{K}(A_1,F_1)$\\
1.\> $F_2\leftarrow A_1\oplus \mathbf{F}_K(F_1)$; \\
2.\> $B_2\leftarrow F_1\oplus \mathbf{F}_K(F_2)$; \\
return $(F_2,B_2)$.
\end{tabbing}
\end{minipage}
\\ \hline
\end{tabular}
}
\end{center}
\end{table*}

Two instantiations of $h$ and $h^{\prime}$, namely using $\poly$ and $\sym{BRW}$  have been considered in~\cite{FAST}. Following the notation of~\cite{FAST},
these instantations are denoted as $\sym{FAST}[\sym{Fx}_m,\poly]$ and $\sym{FAST}[\sym{Fx}_m,\sym{BRW}]$, where $\sym{Fx}_m$ denotes fixed length messages having
$m$ $n$-bit blocks. 

In case of $\sym{FAST}[\sym{Fx}_m,\poly]$, $m\geq 3$ and the hash functions $h,h^{\prime}$ are defined as:
\begin{eqnarray}
h_\tau(T,X_3||\cdots||X_m) &=& \tau\poly_\tau(\mathbf{1},X_3,\ldots,X_m,T); \label{eq:flPoly} \\
h^\prime_\tau(T,X_3||\cdots||X_m) &=& \tau^2\poly_\tau(\mathbf{1},X_3,\ldots,X_m,T). \label{eq:flPoly1}
\end{eqnarray}
In case of $\sym{FAST}[\sym{Fx}_m,\sym{BRW}]$, $m\geq 4$ and $h,h^{\prime}$ are defined as:
\begin{eqnarray}
h_{\tau}(T,X_3||\cdots||X_m) & = & \tau{\sf BRW}_{\tau}(X_3,\ldots,X_m,T); \label{eqn-Fx-brw} \\
h^{\prime}_{\tau}(T,X_3||\cdots||X_m) & = & \tau^2{\sf BRW}_{\tau}(X_3,\ldots,X_m,T). \label{eqn-Fx-brw1}
\end{eqnarray}

\subsection{Description of AEZ \label{subsec-AEZ-desc}}
AEZ~\cite{DBLP:conf/eurocrypt/HoangKR15} was proposed as a candidate for the CAESAR\footnote{https://competitions.cr.yp.to/caesar.html} competition.
The design has several variants which are built from different block ciphers obtained by modifying AES. In the present context, the relevant version is the one where the 
standardised AES algorithm is used.
In AEZ, short messages of lengths less than $2n$ bits are handled differently from the messages whose lengths are at least $2n$ bits. It is the latter 
which is appropriate for disk encryption algorithm and so we consider only this case.
The construction which handles messages of lengths at least $2n$ bits has been called AEZ-Core~\cite{DBLP:conf/eurocrypt/HoangKR15}.
By AEZ, we will mean AEZ-Core where the encryption function is instantiated using the encryption function of AES and we will denote this construction as
AEZ-Core[AES]. A brief description of this algorithm is given below. For complete details, we refer to~\cite{DBLP:conf/eurocrypt/HoangKR15}.

As mentioned in Section~\ref{subsec-FAST-desc}, an $n$-bit string $X$ is identified with an element $X(\alpha)$ of the field $\rF=GF(2^n)$ which is represented using 
$\psi(\alpha)$. The operation $\alpha X(\alpha)\bmod \psi(\alpha)$ is the `multiply by $\alpha$' map and has been called a doubling operation~\cite{Ro04}. 
For $n=128$, the polynomial $\psi(\alpha)=\alpha^{128}\oplus\alpha^7\oplus\alpha^2\oplus\alpha\oplus 1$ is used to construct the field $\rF$.

Given $X\in \{0,1\}^n$, AEZ denotes the doubling operation as $2\cdot X$. This operation can be implemented using bit operations on the string $X$.
Further, for $i\in\mathbb{N}$, AEZ requires the operation $i\cdot X$ which is defined in the following manner.
\begin{eqnarray}\label{eqn-mask}
i\cdot X
& = &
\left\{
\begin{array}{lcl}
\mathbf{0} & \mbox{if} & i=0; \\
X & \mbox{if} & i=1; \\
2\cdot X & \mbox{if} & i=2; \\
2\cdot (j\cdot X) & \mbox{if} & i=2j>2; \\
(2j\cdot X) \oplus X & \mbox{if} & i=2j+1>2. \\
\end{array}
\right.
\end{eqnarray}

Consider a message of length at least $2n$ bits. Write the length as $2nk+\mu$ bits with $0\leq \mu<2n$ and $k\geq 1$. For the disk encryption
application where the sector size is 4096 bytes, since $n=128$, we have $\mu=0$. So, we provide the description of AEZ for the case $\mu=0$.
The number of $n$-bit blocks in the message is $m=2k=2^8$.

Let $\mathfrak{m}$ be such that $m=2\mathfrak{m}+2$. (For $m=2^8$, $\mathfrak{m}=2^7-1$.)
AEZ partitions the message into two parts in the following manner.
The first part consists of $2\mathfrak{m}$ $n$-bit blocks $M_1,M_1^{\prime},\ldots,M_{\mathfrak{m}},M_{\mathfrak{m}}^{\prime}$ and
the second part consists of 2 $n$-bit blocks $M_{\sf x}$ and $M_{\sf y}$.
The ciphertext blocks are $C_i,C_i^{\prime}$, $i=1,\ldots,\mathfrak{m}$ and $C_{\sf x},C_{\sf y}$.

The encryption algorithm of AEZ can be viewed as consisting of three layers. The first and the third layers are built as a sequence of 2-round Feistel
networks. The second layer is essentially a mixing layer. Let $E$ denote the encryption function of AES and for $\beta\in\{0,1\}^n$, define
$\widetilde{E}_K^{i,j}(\beta) = E_K(\beta\oplus (i+1)\cdot I \oplus j\cdot J)$ where $I=E_K(\mathbf{0})$ and $J=E_K(\mathbf{1})$.
The encryption algorithm of AEZ proceeds as follows:
\begin{description}
\item {\em First layer:} for $i=1,\ldots,\mathfrak{m}$,
	$W_i=M_i\oplus \widetilde{E}_K^{1,i}(M_i^{\prime})$;
	$X_i=M_i^{\prime} \oplus \widetilde{E}_K^{0,0}(W_i)$; \\
        $S_{\sf x}=\widetilde{E}_K^{0,1}(M_{\sf y})\oplus M_{\sf x}\oplus X\oplus \Delta$;
        $S_{\sf y}=\widetilde{E}_K^{-1,1}(S_{\sf x})\oplus M_{\sf y}$;
\item {\em Second layer:} for $i=1,\ldots,\mathfrak{m}$,
	$S_i^{\prime} = \widetilde{E}_K^{2,i}(S)$;
	$Y_i = S_i^{\prime}\oplus W_i$; $Z_i = S_i^{\prime}\oplus X_i$;
\item {\em Third layer:} for $i=1,\ldots,\mathfrak{m}$,
	$C_i^{\prime}=Y_i\oplus \widetilde{E}_K^{0,0}(Z_i)$;
	$C_i=Z_i \oplus \widetilde{E}_K^{1,i}(C_i^{\prime})$; \\
	$C_{\sf y}=S_{\sf x}\oplus \widetilde{E}_K^{-1,2}(S_{\sf y})$;
	$C_{\sf x}=S_{\sf y}\oplus \widetilde{E}_K^{0,2}(C_{\sf y})\oplus \Delta\oplus Y$;
\end{description}
Here $X=X_1\oplus\cdots\oplus X_\mathfrak{m}$, $Y=Y_1\oplus\cdots\oplus Y_\mathfrak{m}$, $S=S_{\sf x}\oplus S_{\sf y}$ and $\Delta$ is obtained by processing the tweak.
For the decryption algorithm, we refer to~\cite{DBLP:conf/eurocrypt/HoangKR15}.

In our implementation of AEZ, we have ignored the tweak, i.e., we have taken $\Delta=\mathbf{0}$. Our goal was to compare the implementations of AEZ with FAST. Since the
results indicate that AEZ without tweak is slower than FAST with tweak, it follows that AEZ with tweak will also be slower than FAST with tweak. 

\section{Implementation of \sym{FAST} \label{sec-FAST}}
In this section, we describe the design decisions and the architecture of the two variants of FAST, namely $\sym{FAST}[\sym{Fx}_m,\sym{Horner}]$
and $\sym{FAST}[\sym{Fx}_m,\sym{BRW}]$. The value of $m$ is mentioned below.

The basic design goal was speed and so the implementations were optimised for speed. Nevertheless, we tried to keep the area metric reasonable. The target devices were 
high end fast FPGAs. In particular, we have optimised our designs for the Xilinx Virtex~5 and Virtex~7 families.

As mentioned earlier, we have used the encryption function $E_K$ of AES to instantiate the PRF $\mathbf{F}_K$. So, $n=128$. We have considered 4096-byte disk
sectors, so that the message length is also 4096 bytes which corresponds to 256 128-bit blocks, i.e., $m=256$. 
So, our implementations are those of $\sym{FAST}[\sym{Fx}_{256},\sym{Horner}]$ and $\sym{FAST}[\sym{Fx}_{256},\sym{BRW}]$.
With $m=256$ and a single block tweak, the numbers of blocks in the inputs to the hash functions $h$ and $h^{\prime}$ are both 255. The
255 blocks comprise of 254 blocks arising from $X_3||\cdots||X_{256}$ and one block from the tweak. Since 255 blocks are to be hashed,
for $\sym{FAST}[\sym{Fx}_{256},\sym{Horner}]$, the requirement is to implement 255-block {\poly} while for
$\sym{FAST}[\sym{Fx}_{256},\sym{BRW}]$, the requirement is to implement 255-block \sym{BRW}.

For both $\sym{FAST}[\sym{Fx}_{256},\sym{Horner}]$ and $\sym{FAST}[\sym{Fx}_{256},\sym{BRW}]$, we have implemented two variants,
one with a single core of the AES encryption module and the other with two cores of the AES encryption module. We denote variants
of $\sym{FAST}[\sym{Fx}_{256},\sym{Horner}]$ and $\sym{FAST}[\sym{Fx}_{256},\sym{BRW}]$ using a single AES core as
{\sf FAST}[{\sf AES,Horner}]-1 and {\sf FAST}[{\sf AES,BRW}]-1 respectively.
The variants of $\sym{FAST}[\sym{Fx}_{256},\sym{Horner}]$ and
$\sym{FAST}[\sym{Fx}_{256},\sym{BRW}]$ using two AES cores are denoted as
{\sf FAST}[{\sf AES,Horner}]-2 and {\sf FAST}[{\sf AES,BRW}]-2 respectively.

The two basic building blocks for all of these designs are the encryption function of the AES and a finite field multiplier.

In our implementations, we have used pipelined AES encryption cores, which is most suited for a fast implementation. 
An AES encryption core requires a key generation module. For the
two-core designs the same key generation module is shared by both the cores. 
We consider the AES rounds as pipeline stages, whereas the multiplexers and the XORs at the input of AES have been considered as 
an additional stage, so that the delay of AES 
round is not increased.
As a result, the latency of each AES core is 11 cycles, i.e., the
first block of ciphertext is produced after a delay of 11 cycles and thereafter one cipher block is obtained in each cycle. The
design of the AES cores adopts some interesting ideas reported earlier~\cite{DBLP:conf/africacrypt/BulensSQPR08}. The earlier
design~\cite{DBLP:conf/africacrypt/BulensSQPR08} was that of a sequential AES design tailored for the Virtex 5 family of devices.
An important aspect of this design is that the S-boxes are implemented as $256\times 8$ multiplexers and one S-box fits into 32
six-input LUTs which are available in Virtex 5 FPGAs. We have used the same idea to design the S-boxes of our pipelined AES core.

With $n=128$, the requirement is to compute products in $GF(2^{128})$. For this, we have used a 4-stage pipelined Karatsuba multiplier.
The number of stages was selected to match the maximum frequency of the AES encryption core, which is the only other significant
component in the circuits. The multiplier design is the same as reported in a previous work~\cite{DBLP:journals/jce/ChakrabortyMS18}.

To use the pipelined multiplier efficiently, it is important to schedule the multiplications in such a way that pipeline delays are
minimised. The \sym{BRW} computation is amenable to a very efficient pipelined implementation. This requires identifying an ``optimal''
order of the multiplications so that both pipeline delays and the necessity to store intermediate results are minimised.
A detailed study of such an optimal ordering is available in the literature~\cite{DBLP:journals/tc/ChakrabortyMRS13}. A circuit for computing
\sym{BRW} polynomials on 31 blocks of inputs using a 3-stage pipelined Karatsuba multiplier is
known~\cite{DBLP:journals/tc/ChakrabortyMRS13}. In the present work, the requirement is to compute \sym{BRW} polynomials on 255
blocks using a
4-stage pipelined multiplier. We scale up the earlier design~\cite{DBLP:journals/tc/ChakrabortyMRS13} suitably for our purpose.

For computing \sym{Horner} using a pipelined multiplier the idea of decimation is used. 
 Let $(P_1,P_2,\ldots,P_m)$ and a positive integer $d$ be given.
Let $\chi_i=m-i\pmod d$. The $d$-decimated {\poly} computation~\cite{DBLP:journals/tosc/ChakrabortyGS17} is based on the following
observation.
\begin{eqnarray*}
\lefteqn{{\sf Horner}_\tau(P_1,P_2,\ldots,P_m)} \\
&=& \tau^{\chi_1} {\sf Horner}_{\tau^d}(P_1,P_{1+d},P_{1+2d}, \ldots )
\oplus \cdots
\oplus \tau^{\chi_d} {\sf Horner}_{\tau^d}(P_d,P_{2d},P_{3d}, \ldots ).
\end{eqnarray*}
So, ${\sf Horner}_\tau(P_1,P_2,\ldots,P_m)$ can be computed by evaluating $d$ {\em independent} polynomials at $\tau^d$ and
then combining the results. This representation allows efficient use of a $d$-stage pipelined multiplier, as in each clock, $d$
independent multiplications can be scheduled.

In what follows, we give a detailed description of the architecture of {\sf FAST}[{\sf AES,BRW}]-2 followed by a short description
of the architecture of {\sf FAST}[{\sf AES,Horner}]-2.

\subsubsection{Architecture for {\sf FAST}[{\sf AES,BRW}]-2:}
\begin{figure}
\center
\includegraphics[scale=0.6]{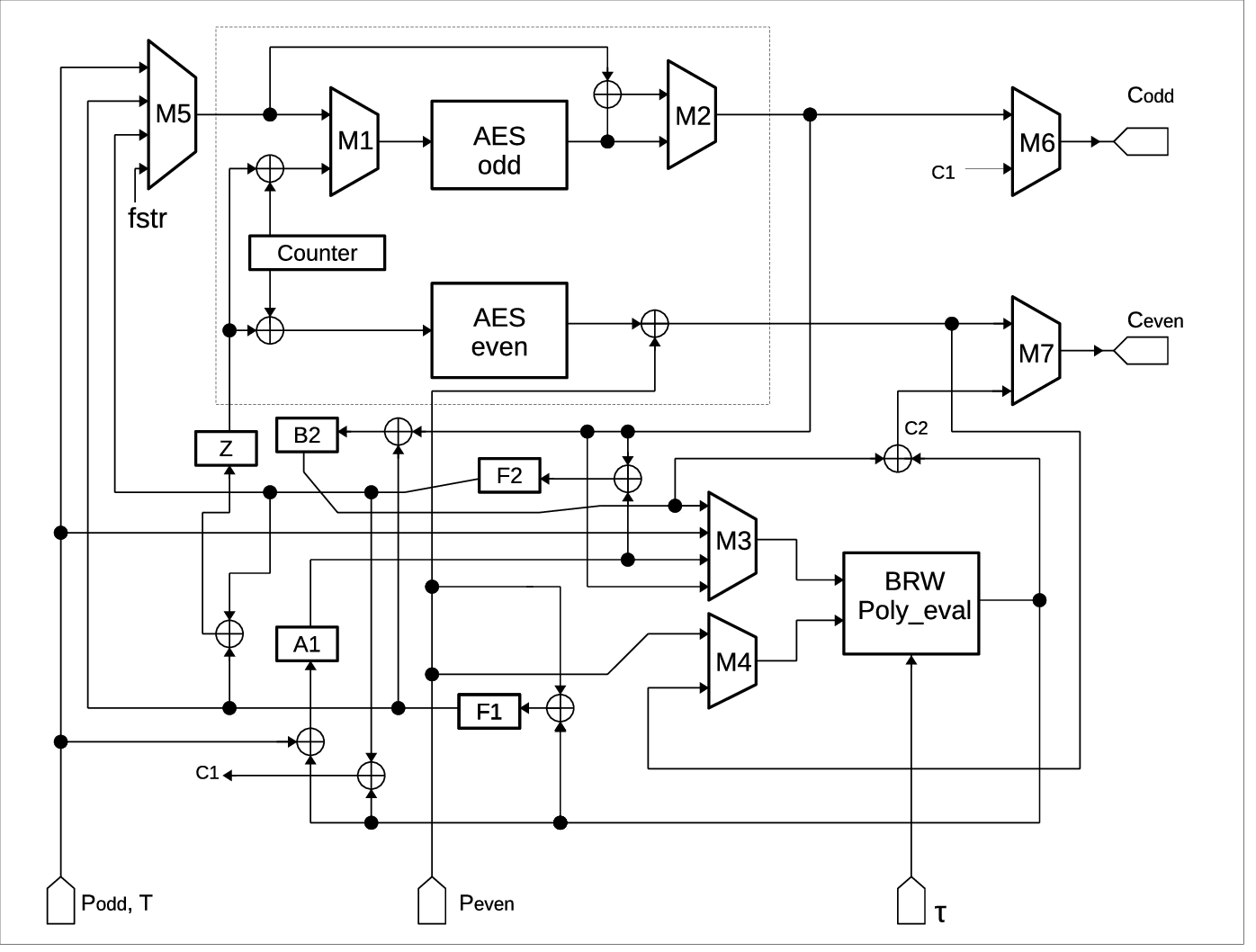}
\caption{\label{fig:db} Architecture for {\sf FAST}[{\sf AES,BRW}]-2.}
\end{figure}

{\sf FAST}[\sym{AES},\sym{BRW}]-2 uses two pipelined AES encryption cores and a 4-stage pipelined multiplier.
An overview of the architecture is shown in Figure \ref{fig:db}. We briefly describe its components and functioning.

The basic components of the architecture are the two AES encryption cores which are denoted as {\bf AESodd} and {\bf AESeven}.
The module for the \sym{BRW} polynomial evaluation using a 4-stage Karatsuba multiplier is shown as {\bf BRWPoly\_eval}.

The two AES cores, two multiplexers {\bf M1} and {\bf M2} and a counter named {\bf Counter} are enclosed inside a dashed
rectangle. This constitutes a module which implements the  counter mode. The module can also perform AES encryption of a single block.
The {\bf AESeven} core is used only in counter mode whereas the {\bf AESodd} core is used for both encryption in the counter mode and
to encrypt single blocks. According to the algorithms in Tables~\ref{tab-encrypt} and~\ref{tab-feistel}, encryption of a single block
is required for the blocks $F_1$ and $F_2$ in the {\sf Feistel} function and for ${\sf fStr}$ in the main function.

The counter has two outputs, one for odd values and the other for even values. The even values are fed directly to the {\bf AESeven}
core and the odd values are fed to the {\bf AESodd} core through the multiplexer {\bf M1}. The block {\bf BRWPoly\_eval} performs the
255-block \sym{BRW} computation. Additionally, this block also computes the single multiplications by $\tau$ required for the
computation of $\mathbf{H}_{\tau}$ and $\mathbf{G}^{\prime}_\tau$ (see Table~\ref{tab-encrypt}).

The registers {\bf Z}, {\bf A1}, {\bf F1}, {\bf F2} and {\bf B2} are used to store the intermediate values and these correspond
to the variables $Z, A_1, F_1, F_2$ and $B_2$ respectively of the algorithms described in Tables~\ref{tab-encrypt} and~\ref{tab-feistel}.

The input ports {\bf Podd} and {\bf Peven} are used to feed in the odd numbered message blocks and even numbered message blocks
respectively. The tweak is also fed in through {\bf Podd}. The hash key is fed in through a separate port. The output ports {\bf Ceven}
and {\bf Codd} output the even and odd numbered cipher blocks respectively.

The multiplexer {\bf M5} selects the input to {\bf AESodd} from one of the four possible inputs, namely, {\bf Podd}, {\bf F1},
{\bf F2} or the string ${\sf fStr}$. The multiplexer ${\bf M1}$ selects either the output of ${\bf M5}$ or $Z \oplus i$, where $i$ is
the output from the odd port of {\bf Counter}. This input design to the ${\bf AESodd}$ core through the multiplexers {\bf M1} and {\bf M5}
allows ${\bf AESodd}$ to encrypt in the counter mode and also to encrypt the required single blocks.

The \sym{BRW} computation module {\bf BRWPoly\_eval} is required to be fed two blocks of plaintext or ciphertext in each cycle. The
multiplexer {\bf M3} provides the first input to {\bf BRWPoly\_eval}. This input is selected by {\bf M3} to be one of {\bf Podd},
{\bf Codd}, {\bf A1} or {\bf B2}. The inputs {\bf Podd} and {\bf Codd} are relevant for \sym{BRW} while the inputs {\bf A1}
and {\bf B2} are relevant when a single-block multiplication is required. The second input to {\bf BRWPoly\_eval} is the output
of the multiplexer {\bf M4} and can be either {\bf Peven} or {\bf Ceven}.

The final outputs of the circuit are selected using multiplexers {\bf M6} and {\bf M7}. Control signals are generated using a
finite state machine which follows the algorithm of \sym{FAST}.

\begin{figure}
\center
\includegraphics[scale=0.6]{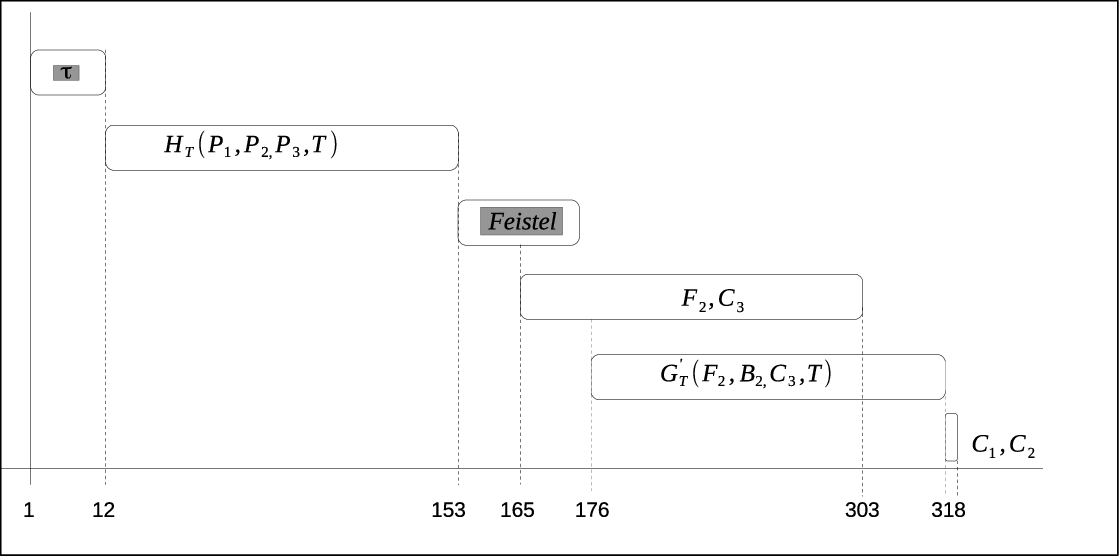}
\caption{\label{fig:time}Time diagram for encryption using {\sf FAST[AES,BRW]}-2.}
\end{figure}

\paragraph{\bf Timing analysis:}
Figure~\ref{fig:time} shows the timing diagram for  {\sf FAST}[\sym{AES},\sym{BRW}]-2. The first 11 clock cycles
are required to compute the hash key $\tau$ by applying the AES encryption module to {\sf fStr}. The computation of the hash function
$\mathbf{H}_{\tau}$ (see~(\ref{eqn-H-Hprime})) requires a 255-block \sym{BRW} computation and two subsequent field multiplications by
$\tau$. The 255-block \sym{BRW} computation requires 127 field multiplications. The 4-stage multiplier has a latency of 4 cycles. So,
the \sym{BRW} computation requires 131 cycles. The two subsequent multiplications require 4 cycles each. The computation of $\mathbf{H}_{\tau}$
is completed after 141 cyles which includes two additional synchronisation cycles.
The Feistel network has two encryptions. The first encryption requires 11 cycles. After the first encryption, both $F_1$ and
$F_2$ are available and so the input $Z=F_1\oplus F_2$ to the counter can be obtained. Let $J_i=Z\oplus \sym{bin}_n(i)$, $i=1,\ldots,254$.
{\bf AESodd} performs the encryptions of $F_1,F_2,J_1,J_3,\ldots,J_{253}$ while {\bf AESeven} performs the encryptions of
$J_2,J_4,\ldots,J_{254}$. {\bf AESodd} and {\bf AESeven} are synchronised such that the encryptions of $J_{2j-1}$ and $J_{2j}$,
$j=1,\ldots,127$, are obtained simultaneously. This allows the computation of $\mathbf{G}_{\tau}^{\prime}$ to start after
the encryptions of $J_1$ and $J_2$ are completed and be executed in parallel with the rest of the encryptions of the counter. The
total computation requires 319 cycles which includes a few synchronisation cycles.

\subsubsection{Architecture for {\sf FAST}[{\sf AES,Horner}]-2:}
To take the advantage of two AES cores in the design of {\sf FAST}[{\sf AES,Horner}]-2 it becomes necessary to use two multipliers.
The reason is the following. The crucial parallelisation is in computing the second hash layer
where the hash of the ciphertexts produced by the counter mode is computed. Since two pipelined AES cores are used to implement the
counter mode, after an initial delay, in each clock cycle two blocks of ciphertexts are produced. So, the hash module has to be
capable of processing two ciphertext blocks in each cycle. For \sym{BRW} based hashing, each multiplication involves two ciphertext blocks.
On the other hand, in the case of {\poly}, each multiplication involves a single block. So, to process two ciphertext blocks in each cycle
it is required to use two multipliers. Each multiplier operates in a 4-stage pipeline. For proper scheduling using the two multipliers,
it is required to use a 8-decimated version of Horner. This allows the scheduling of four independent multiplications to each multiplier
in every clock cycle.

\section{Implementation of AEZ \label{sec-AEZ}}

The design decisions regarding the choice of FPGAs, the block cipher and the message length are the same as that of FAST. In this section, we provide
an overview of the architecture that we have designed to implement AEZ.

\begin{figure}
\includegraphics[width=12cm]{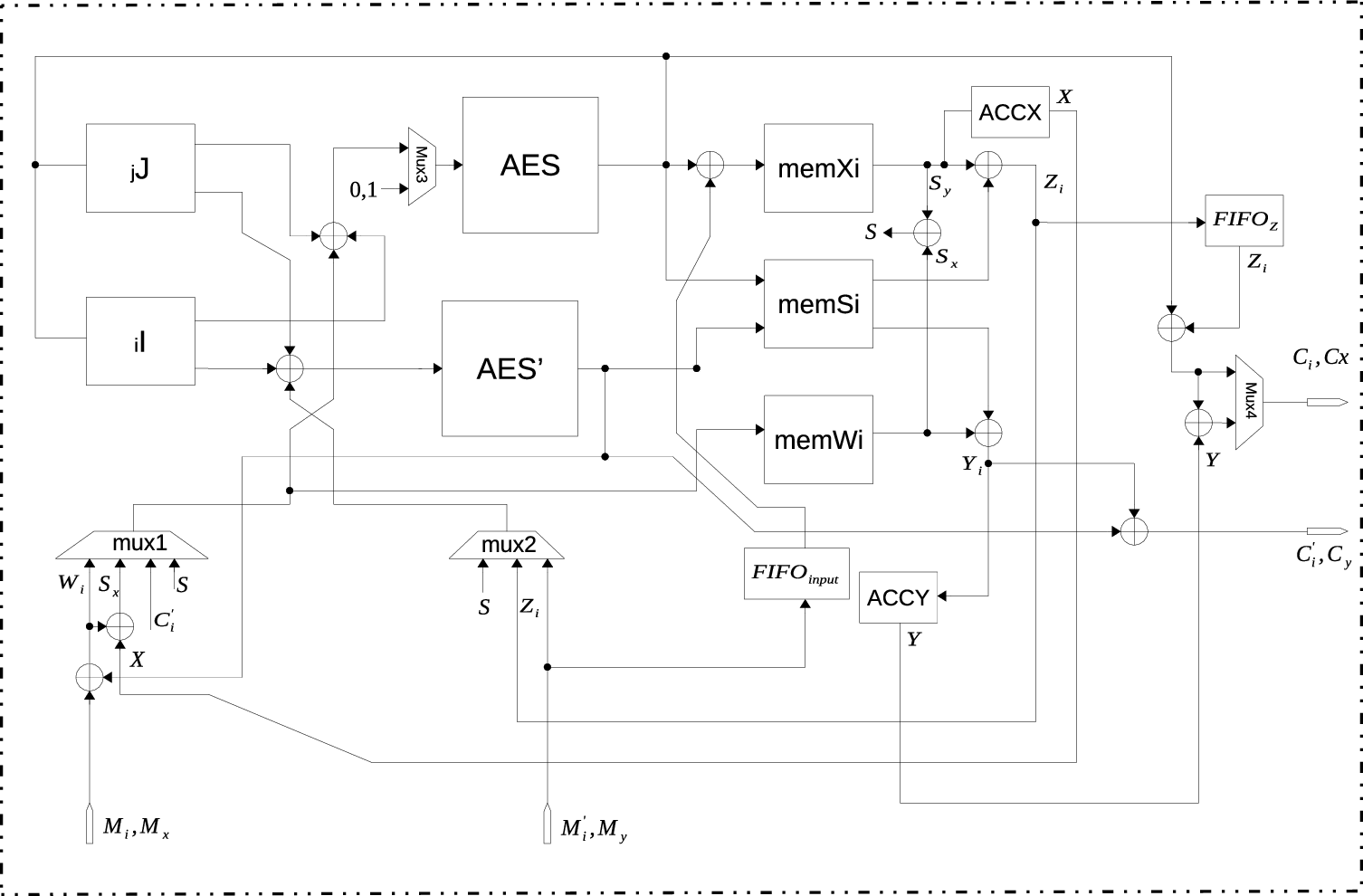}
\caption{\label{fig:aezarch}Pipelined architecture for AEZ using two AES-encryption cores.}
\end{figure}

The architecture in Figure~\ref{fig:aezarch} allows the computation of AEZ encryption/decryption. This implementation
uses two cores. The two AES cores are labeled as
{\bf AES} and {\bf AES}$^{\prime}$ and the inputs to these cores are selected by {\bf mux1} and {\bf mux2} respectively.
For computing the masking values $I$ and $J$ we need the encryptions of $\mathbf{0}$ and $\mathbf{1}$ respectively.
This is enabled using {\bf mux3}. {\bf AES} and {\bf AES}$^{\prime}$ work in parallel to compute $X_i$ and $W_i$ respectively. Since $X_i$
depends on $W_i$, {\bf AES} waits for the first value of $W_i$ to be produced to start the computation. The computed values of
$W_i$ and $X_i$ have to be stored and for that we use the single-port-block-RAMs {\bf memWi} and {\bf memXi} respectively. At a later stage,
the values $S_{\sf x}$ and $S_{\sf y}$ are also stored in {\bf memWi} and {\bf memXi} respectively.
The computation of $X=X_1\oplus\cdots \oplus X_{127}$ is performed by {\bf ACCX}.
The values of $S_i^{\prime}$'s are computed with the two AES cores. Due to data dependencies, these values need to be stored
and for this purpose the dual-port-block-RAM {\bf memSi} is used. In {\bf memSi}, the last value is initialised to $\mathbf{0}$ and so
$Z_{128}=S_{\sf y}\oplus \mathbf{0}=S_{\sf y}$ and $Y_{128} = S_{\sf x}\oplus \mathbf{0}=S_{\sf x}$.
The computation of $Y=Y_1\oplus\cdots\oplus Y_{127}$ is performed by {\bf ACCY}.
The input line marked $C_i^{\prime}$ to {\bf mux1} carries $C_{\sf y}$ at the end.

The masking values $j\cdot J$ can be computed in two different ways.

\begin{figure}
\subfloat[Masks generated on the fly]{
\includegraphics[width=5cm]{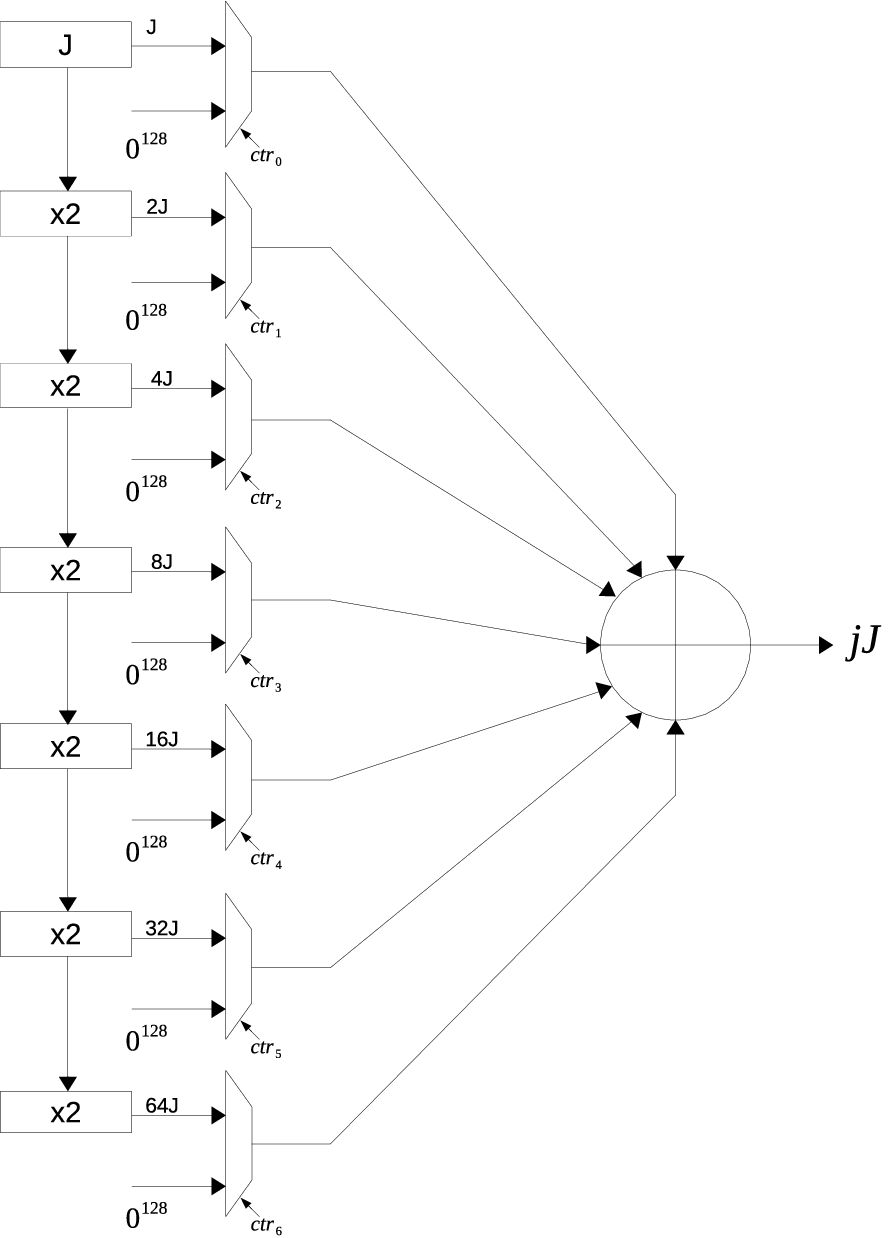}
\label{fig:fly}
}
\subfloat[Precomputing masks using double and add]{
\includegraphics[width=7cm]{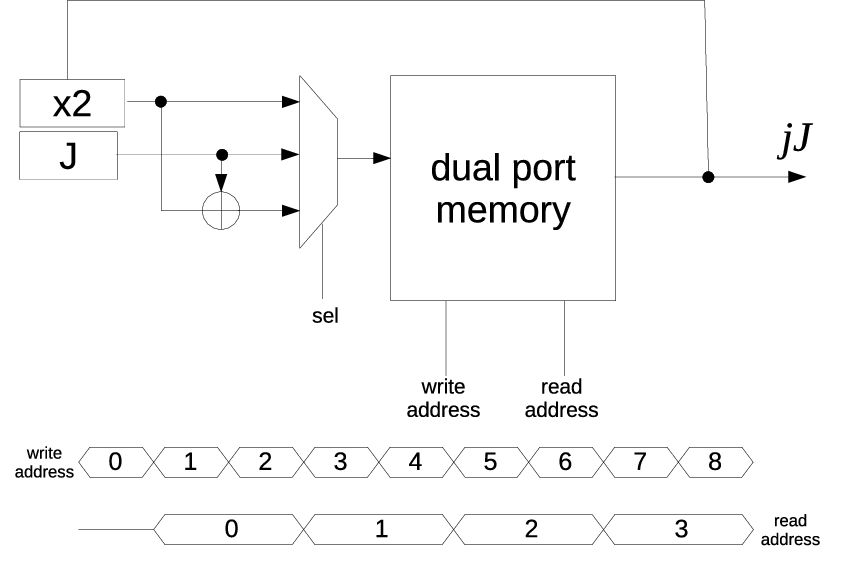}
\label{fig:pre}
}
\vspace{1cm}
\caption{\label{fig:pre_mask}Two ways to compute the masks $j\cdot J$. In the figure $\times 2$ denotes doubling.}
\end{figure}

\paragraph{\bf Pre-computation:}
The values of $I$, $J$ and all the necessary values $i\cdot I$, $j\cdot J$ can be precomputed. Computing $I$ and $J$ take $13$ clock
cycles while $127$ clock cycles are necessary to compute all $j\cdot J$; $i\cdot I$ are only $4$ values and are computed in parallel
with $j\cdot J$.
So the precomputation takes $145$ clock cycles, taking into account the reset time and some clock cycles for synchronisation of the memory.
In Figure~\ref{fig:pre}, we show the architecture to compute all the necessary values of $j\cdot J$ using double and add method. The values
are stored in block RAMs. For the values $i\cdot I$, only the values $\{0,I,2\cdot I,3\cdot I\}$ are required and
so they are computed and stored in registers.

\paragraph{\bf On the fly:} The values $i\cdot I$ are computed as above. For $j\cdot J$ we used the circuit in Figure~\ref{fig:fly}. It
consists of
the computations of $J,2\cdot J,4\cdot J,8\cdot J,16\cdot J,32\cdot J,64\cdot J$. Subsequently, depending on the binary representation of
$j$ some of these are selected and XORed to obtain the required value $j\cdot J$.
For example, $86\cdot J= 64\cdot J \oplus 16\cdot J \oplus 4\cdot J \oplus 2\cdot J$.

\begin{figure}
\includegraphics[width=14cm]{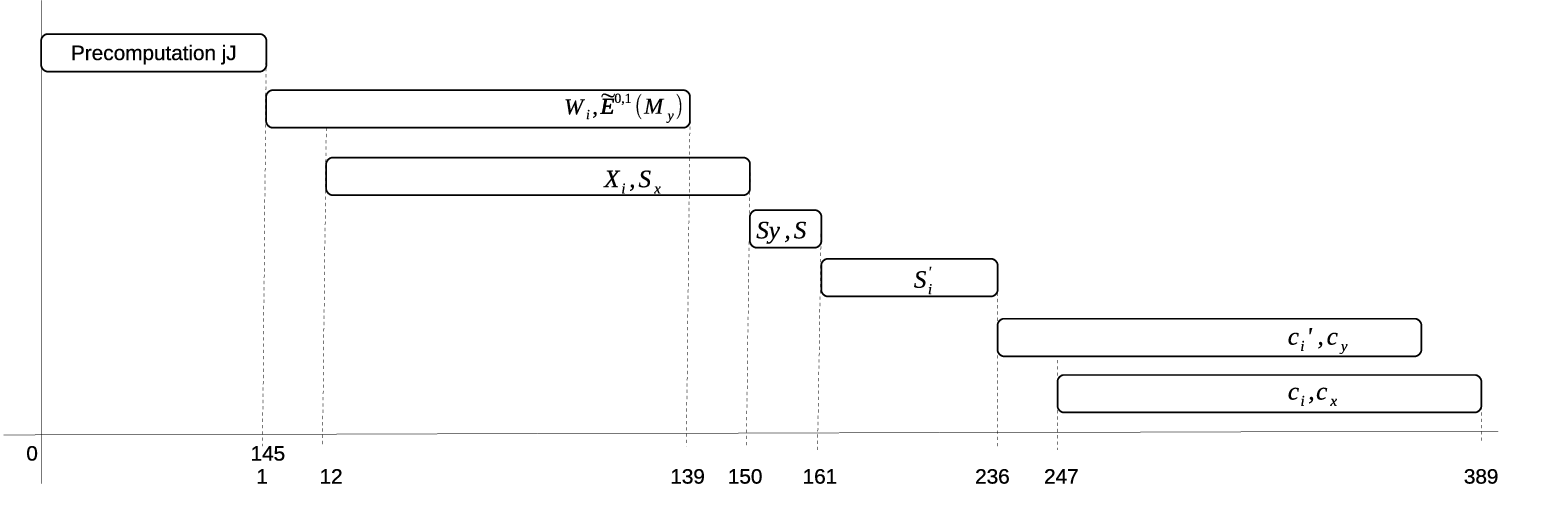}
\caption{\label{fig:aeztime}Timing diagram for encryption using AEZ.}
\end{figure}

The timing diagram for the architecture in Figure~\ref{fig:aezarch} is shown in Figure~\ref{fig:aeztime}. Apart from the
pre-computation, a total of 389 cycles is required to complete the encryption.

The AEZ architecture that we have described uses two AES cores and we name this architecture as AEZ-2. 

\section{Comparative Results \label{sec-compare} }
We present performance data for the implementations of {\sf FAST} mentioned in Section~\ref{sec-FAST} and the implementation of AEZ mentioned in Section~\ref{sec-AEZ}.
The results are compared with the implementations of XCB and EME2. The implementations of XCB and EME2 are taken from~\cite{DBLP:journals/jce/ChakrabortyMS18}. Two 
architectures for each of EME2 and XCB are reported. These are named EME2-1, EME2-2 and XCB-1, XCB-2 respectively.
The hardware resources utilized in these architectures 
along with those used in the different architectures for FAST and AEZ are summarized in Table~\ref{tab-hw}.
\begin{table}
\caption{\label{tab-hw}Summary of the main hardware resources in the architectures of {\sf FAST}, EME2, XCB and AEZ.}
\begin{center}
\setlength{\tabcolsep}{2pt}
\begin{tabular}{|c|c|c|c|c|}\hline \hline
Scheme                         & Pipelined AES   & Pipelined AES   & Sequential AES  & Pipelined  \\
                               & encryption core & decryption core & decryption core & multiplier \\ \hline \hline
{\sf FAST}[{\sf AES,BRW}]-1    & 1               & 0               & 0               & 1           \\ \hline
{\sf FAST}[{\sf AES,Horner}]-1 & 1               & 0               & 0               & 1           \\ \hline
EME2-1                         & 1               & 1               & 0               & 0           \\ \hline
XCB-1                          & 1               & 0               & 1               & 1           \\ \hline \hline
{\sf FAST}[{\sf AES,BRW}]-2    & 2               & 0               & 0               & 1           \\ \hline
{\sf FAST}[{\sf AES,Horner}]-2 & 2               & 0               & 0               & 2           \\ \hline
EME2-2                         & 2               & 2               & 0               & 0           \\ \hline
XCB-2                          & 2               & 0               & 1               & 2           \\ \hline
AEZ-2                          & 2               & 0               & 0               & 0           \\ \hline \hline
\end{tabular}
\end{center}
\end{table}

Some important aspects of the architectures are as follows:
\begin{enumerate}
\item The encryption cores utilised in $\sym{FAST}$ are the same as those utilised in XCB, EME2 and AEZ.
Further, the multiplier utilised in $\sym{FAST}$ is also utilised in XCB.
The sequential decryption core required in XCB was optimised for speed. To match the critical path of the AES encryption core the
sequential decryption core was implemented using T-boxes.
\item EME2 is an encrypt-mask-encrypt type construction which consists of two ECB layers with an intermediate masking.
The ECB layers can be implemented with pipelined AES cores. For decryption, ECB in decryption mode is required;
hence for efficient decryption functionality pipelined AES decryption cores are required to be used.
The second layer of ECB in EME2 can only be computed once the first layer has been completed and so the intermediate results of
the first layer of ECB encryption are required to be stored. Block RAMs are used for this purpose.
\item XCB is a hash-counter-hash type mode which involves a counter mode of operation sandwiched between two polynomial hash layers.
The main encryption/decryption in XCB takes place through a variant of the counter mode (which is different from the counter mode
used in ${\sf FAST}$). The counter mode can be implemented using only the encryption module of AES.
One call to the decryption module of AES is required in XCB for both encryption and decryption. For this, a sequential AES decryption
core is utilised. Thus, XCB-2 uses two pipelined AES encryption cores which does the bulk encryption and in addition uses a sequential
AES decryption core.
\item The polynomial hash layers in XCB consist of ${\poly}$ computations. The second ${\poly}$ computation in XCB can be computed
in parallel with the counter mode. As in case of ${\sf FAST}[{\sf AES,Horner}]$-2 the counter mode in XCB-2 is implemented using two AES
cores. So, in each clock cycle, two blocks of ciphertexts are obtained and to utilise this parallelisation two multipliers are required.
\item For AEZ, we do not consider an architecture consisting of a single AES core. The number of cycles required by such an architecture
will be too high compared to the other schemes.
\item There are two architectures for AEZ, namely, AEZ-2-pre and AEZ-2-otf. In AEZ-2-pre, the required masks are precomputed whereas
in AEZ-2-otf, the required masks are computed on the fly. A total of 145 cycles are required to precompute the masks in
AEZ-2-pre.
\item The architecture for EME2 needs to store intermediate results of lengths equal to the message length. For doing this,
EME2 requires 4 block RAMs. In contrast to EME2, both AEZ-2-pre and AEZ-2-otf require to store more intermediate results
requiring 8 block RAMs. Further, AEZ-2-pre stores the precomputed masks which requires an additional 4 block RAMs. So, overall
AEZ-2-otf requires 8 block RAMs while AEZ-2-pre requires 12 block RAMs.
\end{enumerate}

The performance results presented in Table~\ref{tab:results_v5} are obtained after place and route process in ISE 14.7. The target
device was {\tt xc5vlx330t-2ff1738}. We tried many timing restrictions and the best case is reported.

\begin{table}
\begin{center}
\caption{\label{tab:results_v5}Implementation results for Virtex 5.}
\begin{tabular}{|c|c|c|c|c|c|}
\hline 
Architecture             & \multicolumn{2}{c|}{Area}      & Frequency & Clock   & Throughput \\ \cline{2-3}
                         & slices & blk RAMs                  & (MHz)     & cycles  & (Gbps)     \\ \hline
  AES-PEC                & 2859   & 0                      & 300.56    &  1      & 38.47      \\ \hline
  AES-PDC                & 3110   & 0                      & 239.34    &  1      & 30.72      \\ \hline
  AES-SDC                & 1800   & 0                      & 292.48    &  11     & 3.40       \\ \hline
128-bit mult             & 1650   & 0                      & 298.43    &  1      & 38.20      \\ \hline\hline
{\sf FAST[AES,BRW]}-2    & 7175   & 0                      & 289.56    &  319    & 29.74      \\ \hline
{\sf FAST[AES,Horner]}-2 & 8983   & 0                      & 289.98    &  311    & {\bf 30.55}\\ \hline\hline
XCB-2                    & 9752   & 0                      & 270.52    &  316    & 28.05      \\ \hline
EME2-2                   & 10970  & 4                      & 230.56    &  305    & 24.77      \\ \hline \hline
AEZ-2-pre                & 5646   & 12                     & 269.56    &  389 (+145) & 22.70$\dagger$ \\ \hline
AEZ-2-otf                & 5854   & 8                      & 272.32    &  404    & 22.08      \\ \hline\hline
${\sf FAST[AES,BRW]}$-1  & 5064   & 0                      & 290.57    &  455    & 20.92      \\ \hline
${\sf FAST[AES,Horner]}$-1  & {\bf 4781} & 0               & 291.05    &  565    & 16.88      \\ \hline \hline
   XCB-1                 & 6070   & 0                      & 272.75    &  569    & 15.70      \\ \hline
   EME2-1                & 6500   & 4                      & 233.58    &  561    & 13.64      \\ \hline 
\end{tabular} \\
$\dagger$: ignores the 145 cycles required for pre-computation.

\end{center}
\end{table}

The first part of Table~\ref{tab:results_v5} shows the performance of the basic modules, i.e., the pipelined encryption core (PEC),
the pipelined decryption core (PDC), the sequential decryption core (SDC) and the 128-bit pipelined Karatsuba multiplier.
The decryption cores are not required in \sym{FAST} and AEZ. The pipelined decryption core is required for EME2 and the sequential
decryption core is required for XCB.
The results for individual AES cores in Table~\ref{tab:results_v5} include the area required for the key schedule module.
For the implementations of modes of operation we have implemented only one key schedule, and it is shared between all the AES cores
presented in the architecture.

From the results in Table~\ref{tab:results_v5} we observe the following:
\begin{enumerate}
\item Comparison of area.
\begin{enumerate}
\item AEZ requires two cores but no multiplier and so the number of slices is lesser than those required for 2 core
architectures for $\sym{FAST}$. On the other hand, the number of slices for AEZ is more than the single core architectures for $\sym{FAST}$
which use a single AES core and a multiplier.
\item Of all the two-core architectures, AEZ-2-pre requires the smallest number of slices and the highest number of block RAMs.
{\sf FAST}[{\sf AES, BRW}]-2, on the other hand, requires more slices than AEZ, but no block RAM.
Among the single-core architectures, {\sf FAST}[{\sf AES, Horner}]-1 is the smallest which is also the smallest design overall.
\item In comparison to {\poly}, the module for implementing \sym{BRW} requires more registers and also circuits for squaring.
As a result, {\sf FAST}[{\sf AES, BRW}]-1 requires 283 slices more than {\sf FAST}[{\sf AES, Horner}]-1.
\item For the two-core architectures, {\sf FAST}[{\sf AES, Horner}]-2 requires more area than {\sf FAST}[{\sf AES, BRW}]-2 since
the implementation of {\sf FAST}[{\sf AES, Horner}]-2 requires two multipliers while the implementation of
{\sf FAST}[{\sf AES, BRW}]-2 requires a single multiplier.
\item EME2 is the costliest in terms of area in both categories of single core and double core architectures.
This is because it requires AES decryption cores. Further, both EME2-1 and EME2-2 require four block RAMs in addition to the slices.
\item The overall architecture of XCB is similar to that of {\sf FAST}[{\sf AES, Horner}]. The main difference is that XCB
requires an additional sequential AES decryption core and this results in XCB being costlier than {\sf FAST}[{\sf AES, Horner}]
in terms of area.
\end{enumerate}
\item Comparison of throughput.
\begin{enumerate}
\item Among the two-core architectures, {\sf FAST[AES,Horner]}-2 has the highest throughput while among the single-core
architectures, {\sf FAST[AES,BRW]}-1 has the highest throughput.
\item As computing \sym{BRW} requires about half the number of multiplications required for computing \sym{Horner}, in comparison
to {\sf FAST[AES,Horner]}-1, a significant number of clocks can be saved in computing the first hash in case of {\sf FAST[AES,BRW]}-1.
As a result, the total number of clocks required by {\sf FAST[AES,BRW]}-1 is smaller than that required by {\sf FAST[AES,Horner]}-1
and this leads to a better throughput for {\sf FAST[AES,BRW]}-1.
\item {\sf FAST[AES,Horner]}-2 is marginally better than {\sf FAST[AES,BRW]}-2 in terms of throughput. This is due to the following
reason. {\sf FAST[AES,Horner]}-2 uses two multipliers which compensates for the gain from the use of \sym{BRW} polynomials. Overall,
{\sf FAST[AES,Horner]}-2 requires slightly lesser number of clocks and utilises slightly higher frequency.
\item Both versions of XCB operate at a lower frequency than the corresponding versions of {\sf FAST}. This leads to lower throughput
of XCB compared to {\sf FAST}. The lower frequency of XCB is essentially due to the use of the sequential AES decryption core which
is not present in the architectures for {\sf FAST}.
\item Among the 2-core architectures, AEZ has the lowest throughput while EME2-1 has the lowest throughput overall.
EME2 has the lowest frequency due to the use of the pipelined decryption core, which is absent in all other architectures.
\item The frequency of AEZ is lower than {\sf FAST}. This is due to the use of block RAMs.
\end{enumerate}
\end{enumerate}

To confirm the comparative performance of the different designs, we have also obtained results for the high end Virtex 7 FPGA.
The target device was {\tt xc7vx690t-3fgg1930}. The results are presented in Table~\ref{tab:results_v7}.
Based on Table~\ref{tab:results_v7}, we make the following observations.
\begin{enumerate}
\item The frequency grows significantly in comparison with Virtex 5 results. This is basically a direct effect of the difference of the
fabrication technology between the two families. While Virtex 5 family is built with 65 nm technology, Virtex 7 is built with 28 nm
technology.
\item The number of slices for the AES cores is significantly lesser than the corresponding implementations in Virtex 5. This is due to
the fact that slices in Virtex 7 include 8 Flip-Flops which is 4 more than that in Virtex 5.
\item In some cases, the number of slices grows in comparison with the Virtex 5. Examples are the 128-bit multiplier and
\sym{FAST}[\sym{AES,Horner}]-1. This behaviour can be attributed to the optimisation performed by the tool.
\end{enumerate}

\begin{table}
\begin{center}
\caption{\label{tab:results_v7}Implementation results for Virtex 7.}
\begin{tabular}{|c|c|c|c|c|c|}
\hline 
Architecture             & \multicolumn{2}{c|}{Area}     & Frequency & Clock   & Throughput \\ \cline{2-3}
                         & slices & blk RAMs             & (MHz)     & cycles & (Gbps)      \\   \hline
  AES-PEC                & 2093   & 0                    & 405.02    &  1      & 51.84      \\ \hline
  AES-PDC                & 2352   & 0                    & 352.19    &  1      & 45.08      \\ \hline
  AES-SDC                & 1575   & 0                    & 390.056   &  11     & 4.54       \\ \hline
128-bit mult             & 1884   & 0                    & 404.86    &  1      & 51.82      \\ \hline \hline
 {\sf FAST[AES,BRW]}-2   & 7202   & 0                    & 375.43 &  319    & 38.56         \\ \hline
 {\sf FAST[AES,Horner]}-2 & 8906  & 0                    & 377.03  & 311    & \textbf{39.73}\\ \hline\hline
XCB-2                    & 9330   & 0                    & 358.84  & 316    & 37.21         \\ \hline
EME2-2                   & 11800  & 4                    & 315.58  & 305    & 33.90         \\ \hline \hline
AEZ-2-pre                & 6072   & 12                   & 361.52  & 389 (+ 145)  & 30.45$\dagger$ \\ \hline
AEZ-2-otf                & 5202   & 8                    & 362.78  &  404   & 29.42         \\ \hline\hline
${\sf FAST[AES,BRW]}$-1  & 5024   & 0                    & 377.87  &  455   & 27.21         \\ \hline
${\sf FAST[AES,Horner]}$-1 & \textbf{4783} & 0           & 379.25  & 565    & 21.99         \\ \hline
   XCB-1                 & 5875   & 0                    & 360.67  & 569    & 20.77         \\ \hline
   EME2-1                & 6350   & 4                    & 319.74  & 561    & 18.67         \\ \hline 
\end{tabular} \\
$\dagger$: ignores the 145 cycles required for pre-computation.
\end{center}
\end{table}

\section{Conclusion\label{sec-conclu}}
In this paper, we have presented FPGA implementations of two latest tweakable enciphering schemes, namely FAST and AEZ, geared towards low level disk encryption. 
The implementations have been compared to the IEEE standards XCB and EME2. The results indicate that variants of FAST provide the best throughput and also the
smallest area. These results will be of interest to disk manufacturers and standardisation bodies in adopting new algorithms for deployment.



\end{document}